\documentclass[a4paper,12pt]{article}
\usepackage{amsfonts}
\usepackage[intlimits]{amsmath}
\usepackage[dvips]{graphicx}
\usepackage{psfrag}

\newcommand{\xp}{X^{+}}
\newcommand{\xm}{X^{-}}

\newcommand{\res}[1]{\underset{#1}{\text{res}}}

\begin{document}
\noindent
\thispagestyle{empty}
\vspace{5cm}
\begin{center}
 {\Large\bf Giant magnons in $AdS_4/CFT_3$: dispersion, quantization and finite--size corrections}\\
\vspace{1cm}
Igor Shenderovich\\
\vspace{1cm}
\textit{Saint Petersburg State University, 198504, Ulyanovskaya str., 3, Saint Petersburg, Russia} \\\texttt{shender.i@gmail.com} 
\end{center}

\vspace{1cm}
\begin{abstract}
We study giant magnon solutions in $AdS_4 \times CP^3$. We compute quantum corrections to their dispersion relation. We find out that the one--loop correction vanishes in infinite volume. This implies that the interpolating function $h(\lambda)$ between strong and weak coupling regimes does not have a constant term $\lambda^0$ at strong coupling. We also compute first nonvanishing finite volume correction to the one--loop expression. When compared to the L\"{u}sher formula, our results could provide a nontrivial check of the $AdS_4 \times CP^3$ S--matrix proposed recently in arXiv:0807.1924.
\end{abstract}

\newpage

\section{Introduction}

Nice example of integrable gauge theory is high--energy QCD \cite{Lipatov:1993yb,Faddeev:1994zg,Braun:1998id}. Recently integrability was discovered for $AdS_5 \times S^5$ string theory and gauge theories \cite{Beisert:2005bm,Minahan:2002ve,Kazakov:2004qf,Kazakov:2004nh}. Many new applications of integrability arised together with famous Maldacena's $AdS/CFT$ duality \cite{Maldacena:1997re}.

Recently Aharony, Bergman, Jafferis and Maldacena \cite{Aharony:2008ug} proposed duality of $AdS_4 \times CP^3$ and $CFT_3$ theories. On conformal side of duality they considered a superconformal Chern--Simons theory with $SU(N) \times SU(N)$ symmetry and level $k$. One can find coupling constant in string theory taking the limit $k, N \to \infty$ and holding fixed
\begin{equation}
 \lambda = \dfrac{N}{k} \equiv 8 g^2\;.
\end{equation} 

Itegrability on the string side of the story was shown in~\cite{Stefanski:2008ik,Arutyunov:2008if} in the strong coupling regime, and algebraic curve for string was proposed in~\cite{Gromov:2008bz}. Minahan and Zarembo~\cite{Minahan:2008hf} proposed two--loop Bethe ansatz for the $SU(4)$ sector of $N=6$ Chern--Simons theory, and in the paper of Gromov and Vieira \cite{Gromov:2008qe} all--loop Bethe ansatz was conjectured. Also one should mention works \cite{Chen:2008qq,Nishioka:2008gz,Bak:2008cp}, where both sides of this $AdS_4/CFT_3$ duality were studied. 

Minahan and Zarembo show an integrable spin chain with alternating spins which corresponds to operators in Chern--Simons theory. Elementary excitations of this chain are magnons. Magnons can form bound states \cite{Dorey:2008zy,Dorey:2006dq}. In string theory they correspond to ``dyonic giant magnons''~\cite{Hofman:2006xt}. In fact, they are semi--classical string solutions which one can map to operators with large energy and momenta. For the giant magnons problem of determining spectrum from both sides of duality becomes easier. 

There is an exact dispersion relation for a giant magnon in infinite volume. This relation is believed to be exact for all $\lambda$:
\begin{equation}
 \Delta = \sqrt{\dfrac{Q^2}{4} +4 h^2(\lambda) \sin^2 (p/2)}\;. 
\end{equation} 
where $Q$ is the number of magnons. Asymptotics of $h(\lambda)$ looks like follows: 
\begin{equation}
 h(\lambda) = \begin{cases}
\sqrt{\lambda/2},& \lambda \gg 1\\
\lambda,& \lambda \ll 1
\end{cases}\;.
\end{equation} 

In infinite volume this law can be fixed by symmetry except for the explicit form of $h(\lambda)$ \cite{Beisert:2005tm}. We can evaluate the dispersion relation at one--loop and restrict $h(\lambda)$ at strong coupling. Also we want to compute all spectrum of the theory. So it is highly important to calculate finite volume corrections to the energy.

In this paper we use the algebraic curve technique (see, for example, \cite{Beisert:2005bm,Kazakov:2004qf}) to compute magnon's dispersion law and finite volume corrections the to one--loop expression of ground state energy. The paper is organised as follows. In section \ref{sec:quasimomenta} we quantize the giant magnon (GM) solution using the algebraic curve technology. During this procedure we have to define twists with the help of usual orbifold treatment \cite{Beisert:2005he} . We introduce quasimomenta, which are obtained only from known analytical properties and asymptotics. Matching the asymptotics of quasimomenta we obtain the dispersion law (for infinite volume)
\begin{equation}\label{eq:dispersion}
\Delta_{s} = \sqrt{ \dfrac{Q^2}{4} + 2\lambda \sin^2 \left( \frac{p}{2} \right)} \quad \text{(GM)},\quad  \Delta_{b} = \sqrt{Q^2 + 8 \lambda \sin^2 \left( \frac{p}{4} \right)} \quad \text{(``big'' GM)} \;.
\end{equation} 
This expression coincides with one in \cite{Grignani:2008te}, and it confirms validity of our technique in $AdS_4 \times CP^3$.

Then, in section \ref{sec:quantization} we calculate quantum fluctuations around the classical solutions, perturbing quasimomenta adding the extra poles (for first appearance of this method see \cite{Gromov:2007aq}). In this way we obtain semi--classical spectrum of fluctuations around the GM solution, taking into account different polarizations of excitations. It occurs that all fluctuation energies are given by the same function.

In section \ref{sec:one_loop_corr} we discuss ground state energy around classical solutions, i.e. one--loop energy shift. Like in \cite{SchaferNameki:2006gk} and \cite{Gromov:2007cd} (see also \cite{SchaferNameki:2006jf} for similar methods) it is useful to rewrite it as contour integral, which can be calculate in GM case. Using the saddle--point method, we obtain asymptotics in powers of $\sqrt{\lambda}$. We hope that this computation will be extremely helpful for checking S--matrix recently obtained in \cite{Ahn:2008aa}.

\textbf{Note added}. While this paper was in preparation we received the paper \cite{Lee:2008ui} with some overlapping results concerning giant magnon classical solutions. 

\section{Magnons: algebraic curve, quasimomenta and dispersion}\label{sec:quasimomenta}

GM was introduced in \cite{Hofman:2006xt} in the context of gauge/string correspondence. Precisely, they suggested the relation between the spin chain magnon states and specific rotating semiclassical string states on $\mathbb{R} \times S^2$. In $AdS_4/CFT_3$ theory it was discussed in \cite{Grignani:2008is} (for similar computation in $AdS_5/CFT_4$ see \cite{Arutyunov:2006gs}). It is a classical soliton solution on worldsheet, which corresponds to the $Q=1$ fundamental excitation of the gauge theory. One should mention that there are two types of GM (see \cite{Gaiotto:2008cg}). First type magnon lies in $CP^1 \approx S^2$ space, second type --- in $RP^2 \subset CP^3$. We should call GM as ``big'' GM, if it consists of two arbitrary GM's. We have a condition that verify our dispersion \eqref{eq:dispersion}:
\begin{equation}
 \Delta_b (P) = 2 \Delta_s (p) = 2 \Delta_s (P/2),
\end{equation}
where $P$ and $p$ --- momenta of ``big'' GM and small GM consequently. The infinite volume dispersion is exact, non--relativistic and looks like in \eqref{eq:dispersion}, where $p$ is the magnon momentum. For consequences o the symmetry see for example \cite{Beisert:2005tm}.
\subsection{Giant magnon}
The algebraic curve \cite{Gromov:2008bz} formalism maps solutions in $AdS_4 \times CP^3$ to a set of quasimomenta $\{q_1(x) \ldots q_{10}(x)\}$. This map looks like follows. Consider a diagonalization of monodromy matrix
\begin{equation}
 \Omega (x) = P \exp \int d\sigma J_{\sigma}(x),
\end{equation} 
where $J(x)$ is a flat connection on equations of motion. Exactly, 
\begin{equation}
 J(x) = \dfrac{j+x \ast j}{x^2-1}\;.
\end{equation} 
Here $j$ encapsulates equations of motion for fields in a very nice form:
\begin{equation}
 d \ast j =0\;.
\end{equation}
One can see that $\Omega(x)$ is analytic on $\mathbb{C}$ except $x=\pm 1$. The eigenvalues of this matrix (i.e. quasimomenta) form Riemann surface (with poles). In our case the logarithms of these quasimomenta can be organised in Riemann surface with 10 branches. These quasimomenta are not independent:
\begin{equation}
 \{q_1,q_2,q_3,q_4,q_5\} = -\{q_{10},q_9,q_8,q_7,q_6\}\;.
\end{equation} 

\begin{center}
\begin{figure}
\centering \psfrag{q3}{\Large $q_3$} \psfrag{q4}{\Large $q_4$} \psfrag{q5}{\Large $q_5$} \psfrag{q6}{\Large $q_6$} \psfrag{q7}{\Large $q_7$} \psfrag{q8}{\Large $q_8$} \psfrag{e1}{\Large 1}\includegraphics[width=5cm]{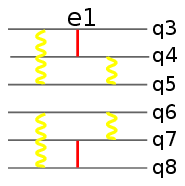} \hspace{1cm} \psfrag{q3}{\Large $q_3$} \psfrag{q4}{\Large $q_4$} \psfrag{q5}{\Large $q_5$} \psfrag{q6}{\Large $q_6$} \psfrag{q7}{\Large $q_7$} \psfrag{q8}{\Large $q_8$} \psfrag{e1}{\Large 1} \includegraphics[width=5cm]{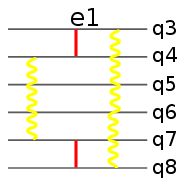}
\caption{Left picture corresponds to usual GM and the right one --- to a ``big'' GM solution. Red lines correspond to a pole, yellow --- to a log cut.}\label{fig:alg_curve}
\end{figure}
\end{center}

Let us describe the analytic structure of quasimomenta more precisely. The algebraic curve for magnon solution looks as shown in Fugure \ref{fig:alg_curve}.

In terms of variables $\xp, \xm$ (in fact, they are conjugated), the magnon momentum is 

\begin{equation}
 p=\dfrac{1}{i} \log \dfrac{\xp}{\xm},
\end{equation}
and it is constrained by the charge
\begin{equation}
 Q= \dfrac{2g}{i} \left( \xp + \dfrac{1}{\xp} - \xm - \dfrac{1}{\xm} \right)\;.
\end{equation}
As in \cite{Minahan:2006bd}, we can work with the log cut description of the GM solution. We have an inversion symmetry for quasimomenta:
\begin{equation}
\begin{array}{cc}
q_1(1/x) = - q_1(x), & q_3(1/x) = - q_4(x) , \\
q_2(1/x) = - q_2(x), & q_5(1/x) =   q_5(x)\;.\\
\end{array}
\end{equation}  
and some asymptotic rules:
\begin{eqnarray}
\left( \begin{array}{l}
q_1 \\
q_2 \\
q_3 \\
q_4 \\
q_5  \\
\end{array}\right) \underset{x\to \infty}{\to} \frac{1}{2gx} \left(
\begin{array}{l}
L + E +S \\
L + E -S \\
L - M_r\\
L + M_r - M_u - M_v\\
M_v - M-u
\end{array}\right), \quad
\left( \begin{array}{l}
q_1 \\
q_2 \\
q_3 \\
q_4 \\
q_5  \\
\end{array}\right)  \underset{x\to \pm 1}{\to} \frac{1}{2(x\pm1)} \left(
\begin{array}{l}
\alpha \\
\alpha \\
\alpha\\
\alpha\\
0
\end{array}\right)\;.
\end{eqnarray}
where $M_u$, $M_v$, $M_r$ are quantum numbers characterizing the $SU(4)$ state, belongs to representation with Dynkin labels $(L-2M_u + M_r,M_u+M_v-2M_r,L-2M_v+M_r)$. It turns out that in Bethe Ansatz Equations $Q$ corresponds to number of roots in one Bethe string. The density of these Bethe roots is roughly constant and it introduces logarithmic cut on Riemann surface (see for more details \cite{Dorey:2008zy,Dorey:2006dq}). These rules fix the charges uniquely:
\begin{equation}
\begin{array}{l}
q_1(x)= -q_{10}(x) = \dfrac{\alpha x}{x^2-1} \,\\[0.3cm]
q_2(x)= -q_9(x) = \dfrac{\alpha x}{x^2-1} \,\\[0.3cm]
q_3(x)= -q_8(x) = \dfrac{\alpha x}{x^2-1} + \dfrac{1}{ i} \log\dfrac{x  - 1/\xm}{x - 1/\xp} + \phi_1 \,\\[0.3cm]
q_4(x)= -q_7(x) = \dfrac{\alpha x}{x^2-1} + \dfrac{1}{ i} \log\dfrac{x - \xp}{x - \xm } + \phi_1\,\\[0.3cm]
q_5(x)= -q_6(x) = \dfrac{1}{i}  \left( \log\dfrac{x-\xp}{x-\xm} - \log\dfrac{x - 1/\xm}{x - 1/\xp} \right) \;.
\end{array}
\end{equation}
In these equations $\phi_1$ is a twist introduced in \cite{Gromov:2007ky} (see \appendixname\ \ref{twists}  for computation details).  It should be equal to $-p/2$ to satisfy boundary conditions. One can check that asymptotics of $q_5(x)$ as $x \to \infty$ is as it should:
\begin{equation}
 q_5(x) =  - \dfrac{Q}{2gx} + O\left(\dfrac{1}{x^2}\right)\;.
\end{equation}

Asymptotics of quasimomenta shouldn't contain energy. So we can extract the dispersion law from this condition. Let us write down the asymptotics for $q_3(x)$ and $q_4(x)$ (asympotics for $q_1(x)$ and $q_2(x)$ are rather trivial):
\begin{equation}
\begin{array}{l}
 q_3(x)= \dfrac{\alpha + \frac{i}{\xm} - \frac{i }{\xp}}{x} + O\left( \dfrac{1}{x^2} \right), \, \\[0.3cm]
 q_4(x)=  \dfrac{\alpha - i (\xm - \xp)}{x} + O\left( \dfrac{1}{x^2} \right)\;.
\end{array}
\end{equation}

In our case $2g \alpha = E + L=\Delta_s - Q/2$, so we obtain \eqref{eq:dispersion}. Let us note that this is in perfect agree with \cite{Grignani:2008is} at strong coupling. Moreover, we obtain expression for the dyonic GM with arbitrary $Q$. For $Q=1$ dyonic magnon reduces to the simple GM from introduction. 

\subsection{``Big'' giant magnon}\label{sec:def_big_GM}
Here we have $p=\dfrac{2}{i}\log\dfrac{X^+}{X^-}$ because there are two type of excitations ($u$ and $v$). The set of charges looks as follows (see Figure \ref{fig:alg_curve}):
\begin{equation}
\begin{array}{l}
q_1(x)= -q_{10}(x) = \dfrac{\alpha x}{x^2-1} \,\\[0.3cm]
q_2(x)= -q_9(x) = \dfrac{\alpha x}{x^2-1} \,\\[0.3cm]
q_3(x)= q_4(x) = -q_7(x) = -q_8(x) = \dfrac{\alpha x}{x^2-1}+\dfrac{1}{i} \log \dfrac{x-\xp}{x-\xm}+\dfrac{1}{i} \log \dfrac{x-1/\xm}{x-1/\xp} + \phi_2\,\\[0.3cm]
q_5(x)= -q_6(x) = 0 \;,
\end{array}
\end{equation} 
where $\phi_2$ is the twist for this case. As before let us write down the asymptotics for $q_3$  (asympotics for $q_1$ and $q_2$ are trivial and $q_4=q_3$):
\begin{equation}
 q_3(x)= \dfrac{\alpha -\frac{1}{i}  \left( \frac{1}{\xm} - \frac{1 }{\xp}-\xm+\xp  \right)  }{x} + O\left( \dfrac{1}{x^2} \right), 
\end{equation} 
If $Q$ is the number of roots of type $u$ we expect $q_1,q_2\simeq (L+E)/(2gx)$, $q_5\simeq 0$ and $q_4=q_3\simeq (L-Q)/(2gx)$. 
Thus $2g \alpha = E +L$, and therefore we obtain $\Delta_b$ as in \eqref{eq:dispersion}.

One should notice that the dispersion law for usual GM differs from the one for the ``big'' GM. 

\subsection{Local charges}\label{sec:local_charges}

We know the explicit expression for quasimomenta so we can compute local conserved charges. As mentioned in \cite{Gromov:2008qe}, 
\begin{equation}
 q_1+q_2-q_3-q_4 = \mathcal{Q}_1^s+ G_4+G_{\bar 4} -\bar G_4 - \bar G_{\bar 4},
\end{equation} 
where bar means $1/x$ instead of $x$ in the argument. $G_4+G_{\bar 4}$ is generating function for all conserved charges. Let us consider the small magnon case. Computation gives 
\begin{equation}
 G_4+G_{\bar 4} = \dfrac{1}{i}\log \left( \dfrac{1/x-XP}{1/x-XM} \right) = - \sum_{n=0} \mathcal{Q}_{n+1}^s x^n,
\end{equation} 
where 
\begin{equation}\label{eq:conserved_charges}
\mathcal{Q}_{n+1}^s = \dfrac{i}{n+1} \left(\left( \dfrac{1}{\xp}\right)^{n+1} - \left( \dfrac{1}{\xm}\right)^{n+1} \right) 
\end{equation} 

For the ``big'' GM one can check that $\mathcal{Q}_{n+1}^b = 2 \mathcal{Q}_{n+1}^s$. 

\section{GM quantization}\label{sec:quantization}

Now we will quantize the classical solutions from section \ref{sec:quasimomenta}. We will calculate explicitly only the ``big'' GM case, for the usual GM the procedure is the same. So let us perturbe the quasimomenta, introduced in subsection (\ref{sec:def_big_GM}) by $\delta q$. One should notice that there are a number of different polarizations of excitations $N_{ij}$, like in \cite{Gromov:2008ie}:
\begin{equation}
 \delta q_i (x)  \sim \eta_i N_{ij}^n \dfrac{\alpha (x^{ij}_n)}{x-x^{ij}_n}, 
\end{equation} 
where $\eta_i$ are signs of the residues and 
\begin{equation}
 \alpha (x) = \dfrac{x^2}{2g(x^2-1)}.
\end{equation} 

Different possible choices of charges correspond to different string polarizations:

\begin{eqnarray}
\nonumber
CP^3&:&   \qquad (3,5), (3,6), (3,7), (4,5), (4,6)\\
\nonumber
AdS&:& \qquad (1,9), (2,9), (1,10)\\
\text{Fermions} & :& \qquad (1,5), (1,6), (2,5), (2,6); (1,7), (2,7), (1,8), (2,8)
\end{eqnarray}

When adding the poles we should satisfy several conditions: 

\begin{itemize}
\item fluctuations $\delta q_i(x)$ can have poles at $x=\pm 1$, but those must be syncronized on different sheets of Riemann surface;
\item final expression should satisfy $x \to 1/x$ symmetry.
\item there should be a ``feedback'' of original solutions to inserting these fluctuations, so we have to shift the log cut. Close to $X^{\pm}$, the quasimomenta should behave as
\begin{equation}
 \delta q (x) \sim \partial \log (x- X^{\pm}) = \dfrac{1}{x- X^{\pm}}\;.
\end{equation} 
\end{itemize}

For example $q_5(x)=0$, so in $\delta q_5(x)$ we should add only poles at $x_n$ and $1/x_n$ (for symmetry $x\to 1/x$). In $q_3(x)$ and $q_4(x)$ we have logs with branch points at $X^{\pm}$ and $1/X^{\pm}$, so we have to add these points as poles in $\delta q_{3,4}(x)$. 

These analytic conditions fix the asymptotic, which on the other hand is fixed by the global charges of the theory. Let us write down only half of charges, because other half is symmetric:

\begin{eqnarray}
\delta \left( \begin{array}{l}
q_1 \\
q_2 \\
q_3 \\
q_4 \\
q_5  \\
\end{array}\right) & \simeq  & \frac{1}{2gx} \left(
\begin{array}{ccccc|ccc}
\delta E & & & & & & N_{19} & 2N_{1,10}\\
\delta E & & & & & 2N_{29} & N_{19} &\\
& & -N_{35} & -N_{36} & -N_{37} & & & \\
-N_{45} & -N_{46} & & & -N_{37} & & &\\
N_{45} & -N_{46} & N_{35} & -N_{36} & & & &  \\
\end{array}\right) + \nonumber \\
&+& \frac{1}{2gx}\left( \begin{array}{cccc|cccc}
  &   & N_{15} & N_{16} &   & N_{17} &   & N_{18}\\
N_{25} & N_{26} &   &   & N_{27} &   & N_{28} &   \\
  &   &   &   &   &   & -N_{28} & -N_{18} \\
  &   &   &   & -N_{27} & -N_{17} &   &   \\
N_{25} & -N_{26} & N_{15} & -N_{16} &   &  &   &  
\end{array}\right) \equiv \frac{1}{2gx}V
\end{eqnarray}

As we know, $\delta q$ has only poles (with known residues), so we can write down their expicit expression
\begin{eqnarray}
&&\delta q(x) = V_0 \dfrac{\alpha(x)}{x-x_n} +M V_0 \dfrac{\alpha(1/x)}{1/x-x_n} + \left(\begin{array}{c}
 1\\
1\\
0\\
0\\
0\\
\end{array}\right) \dfrac{\delta E x}{2g(x^2-1)}  + \left(\begin{array}{c}
0\\
0\\
1\\
1\\
0\\
\end{array}\right) \dfrac{A^{+}\alpha(x)/x}{x-\xp} + \nonumber\\
&&+ \left(\begin{array}{c}
0\\
0\\
1\\
1\\
0\\
\end{array}\right) \dfrac{A^{-}\alpha(x)/x}{x-\xm} + M\left(\begin{array}{c}
0\\
0\\
1\\
1\\
0\\
\end{array}\right) \dfrac{A^{+}\alpha(1/x)x}{1/x-\xp} + M\left(\begin{array}{c}
0\\
0\\
1\\
1\\
0\\
\end{array}\right) \dfrac{A^{-}\alpha(1/x)x}{1/x-\xm},
\end{eqnarray} 
where $V_0$ means $V$ with $\delta E = 0$ (because we want to find it from the explicit expression) and $M$ is a matrix which connect $\delta q(x)$ and $\delta q(1/x)$: $\delta q(x) = M \delta q (1/x)$. 

We need to match poles on the different sheets of Riemann surface. $\delta q_1(x)$ and $\delta q_2(x)$, $\delta q_3(x)$ and $\delta q_4(x)$ have matching poles by pairs, so we need to fit the poles between $\delta q_2(x)$ and $\delta q_3(x)$:
\begin{equation}
\res{x=\pm 1} \left( \delta q_2(x) - \delta q_3(x) \right )=0 \;.
\end{equation} 

Due to the symmetry $x\to 1/x$ we should request $\delta q(0) = 0$ or:
\begin{equation}
 \dfrac{A^{+}}{\xp} + \dfrac{A^{-}}{\xm} = 0.
\end{equation} 

Solving these equations we obtain $\delta E$:
\begin{equation}
\label{eq:delta_E}
 \delta E = \sum_{i,j \in E_1} N_{ij}^n \Omega_{ij}^n + 2\sum_{i,j \in E_2}N_{ij}^n \Omega_{ij}^n,
\end{equation} 
and $\Omega(x)$, $E_1$, $E_2$ are given by
\begin{equation}
\begin{array}{rl}
\Omega_{ij}(x) &= \dfrac{1}{x^2-1} \left( 1- \dfrac{\xp + \xm}{\xp \xm +1} x \right),\\[0.5cm]
E_{1} &= \{(1,5),(1,6),(2,5),(2,6),(3,5),(3,6),(4,5),(4,6)\},\\
E_{2}&= \{(1,7),(1,8),(1,9),(1,10),(2,7),(2,8),(2,9),(3,7)\}\;.
\end{array}
\end{equation} 

One should mention that all fluctuations are given by the same fucntion (as it was in \cite{Gromov:2008ie}), i.e. $\Omega_n^{ij} = \Omega (x_n^{ij})$ for all ($ij$). But the map $x\leftrightarrow n$ is given by the charges discontinuity 
\begin{equation}
 q_i(x_n) - q_j (x_n) = 2 \pi n, 
\end{equation} 
so fluctuations are not all the same. 

Structure of perturbation is different for usual GM. In particular, there will be poles at $X^{\pm}$ and at $1/X^{\pm}$ in $q_5(x)$; $q_4(x)$ must have poles at $X^{\pm}$ and $q_3(x)$ --- at $1/X^{\pm}$ (there are also obtained from the shifting the log cut). However, $\Omega (x)$ remains exactly the same.

\section{Finite size GM: one-loop corrections}\label{sec:one_loop_corr}

Let us study the one-loop energy corrections, i.e. the ground state energy around a classical solution:
\begin{equation}
 \delta \epsilon_{1-loop} = \dfrac{1}{2} \sum_{n \in \mathbb{Z}} \sum_{ij} (-1)^{F_{ij}} \Omega_{ij}\;.
\end{equation} 
This sum is graded, i.e. $F_{ij}=0$ in case of fermion excitation and $F_{ij}=1$ if bosonic one.
Further calculation of $\delta \epsilon_{1-loop}$ goes along one in \cite{Gromov:2008ie} and so we have
\begin{equation}
\delta \epsilon_{1-loop} = - \sum_{ij} \gamma_{ij} (-1)^{F_{ij}} \oint_{\mathbb U} \dfrac{dx}{4i}\dfrac{q'_i-q'_j}{2\pi}\cot \left( \dfrac{q_i - q_j}{2} \right) \Omega(x),
\end{equation} 
where $\gamma_{ij} = 1$ or 2, according to coefficients in \eqref{eq:delta_E}. Expanding the cotangent when the quasimomenta are large we obtain
\begin{equation}
 \cot \left( \dfrac{q_i-q_j}{2} \right) = \pm i (1 + 2 e^{\pm i (q_i - q_j)}+ \ldots ) \;.
\end{equation} 
and using that 
\begin{equation}
 \sum_{ij} \gamma_{ij} (-1)^{F_{ij}} (q'_i - q'_j) =0,
\end{equation}  
(similar to result in \cite{Minahan:2006bd} and \cite{Gromov:2008ie}) we can prove the cancellation of the leading term in the one--loop shift, so we have to calculate the leading correction, which corresponds to second term in the cotangent expansion. After a little algebra, we can find
\begin{equation}\label{eq:one_loop_correction_def}
 \delta \epsilon_{1-loop} = \oint_{\mathbb U} \dfrac{dx}{2\pi i } \partial_x \Omega(x) \sum_{ij} (-1)^{F_{ij}} e^{-i (q_i - q_j)} \;.
\end{equation} 

While taking the of sum in this expression one should notice that all the terms without $q_5$ in exponent would be subleading to others. So let us calculate only terms with $q_5$. We consider a regime with $Q \ll g$ so for a giant magnon we have $1/\xp = \xm$. In this approximation we obtain
\begin{equation}\label{eq:graded_sum_GM}
 \sum_{ij} (-1)^{F_{ij}} e^{-i (q_i - q_j)} = 4 \exp \left( -\dfrac{ix\alpha }{x^2-1} \right) \dfrac{(x+1)(1-\xm )}{x \xm -1}\;.
\end{equation} 

Plugging this expression in the integral \eqref{eq:one_loop_correction_def} and using the saddle--point method, we can write down asymptotics of the first quantum correction:
\begin{equation}\label{eq:one_loop_GM}
 \delta \epsilon_{1-loop} = \dfrac{8e^{-\alpha/2} (1 - \sec(p/2))}{\sqrt{\pi \alpha}} + O \left( \dfrac{1}{\alpha} \right) = \dfrac{8\sqrt{\frac{2g}{E+L}} e^{-\frac{E+L}{4g}} (1 - \sec(p/2))}{\sqrt{\pi}} + O \left( \dfrac{g}{E+L} \right) \;.
\end{equation} 

For a ``big'' GM calculations are similar, except for the sum in \eqref{eq:graded_sum_GM}. One can obtain 
\begin{equation}
 \sum_{ij} (-1)^{F_{ij}} e^{-i (q_i - q_j)} =4\exp \left( -\dfrac{ix\alpha }{x^2-1} \right) \dfrac{(1-x^2)((\xm)^2-1)}{(x \xm -1)^2}\;,
\end{equation} 
and the one--loop correction is 
\begin{equation}\label{eq:one_loop_big_GM}
 \delta \epsilon_{1-loop}  = -\dfrac{16\sqrt{\frac{2g}{E+L}} e^{-\frac{E+L}{4g}} \tan ^2 p}{\sqrt{\pi}} + O \left( \dfrac{g}{E+L} \right)\;.
\end{equation} 

\section{Conclusion}

We construct the algebraic curve for two classical solutions in $AdS_4 \times CP^3$, the so called giant magnons. There are two type of magnons in this theory, which live in different subspaces of $CP^3$. Using algebraic curve technique we identify the dispersion \eqref{eq:dispersion} relations and all local conserved charges \eqref{eq:conserved_charges} of the giant dyonic magnons.

Algebraic curve also allows to find out spectrum of quantum fluctuation and compute one--loop shift of all local conserved charges and in particular to the energy. Using this techinque we find giant magnon excitations in $AdS_4 \times CP^3$. The results for $AdS_4 \times CP^3$ are quite similar to the ones in $AdS_5 \times S^5$, but nevertheless they have remarkable difference (see \eqref{eq:delta_E} and remark in the end of section~\ref{sec:quantization}). 

Also we have obtained the one--loop finite-size correction to the magnon excitation. It turns out that these corrections differ from one in $AdS_5 \times S^5$. Corrections for usual GM and ``big'' GM are different (see \eqref{eq:one_loop_GM} and \eqref{eq:one_loop_big_GM}). These expressions can be useful for checking S--matrix, as it was done in $AdS_5 \times S^5$ through obtaining L\"{u}sher F--term. In this case we had (see \cite{Gromov:2008ie}):
\begin{equation}
\delta E_a^F = \oint \dfrac{dx}{2\pi i}\partial_x \Omega(x) e^{-\frac{4\pi i Jx}{\sqrt{\lambda}(x^2-1)}} \sum_b (-1)^{F_b} S_{ba}^{ba} (q^{\star} (q), p)\;,
\end{equation} 
with $S$--matrix in the right hand side. So we could check our computations of the one--loop shifting in $AdS_5 \times S_5$. It will be interesting to investigate such possibility in $AdS_4 \times CP^3$. 

\section{Acknowledgments}

This work was supported by the ``Dynasty'' foundation. I would like to thank V.~Mikhaylov, D.~Serban, P.~Vieira and especially N.~Gromov for helpful discussions. Also I am very grateful to organizers of Les Houches Summer School, where work was done, for hospitability.

\appendix
\section{Twists computation}\label{twists}

Let us rewrite the explicit magnon solution from \cite{Grignani:2008te}:
\begin{equation}
 Z = \dfrac{1}{\sqrt{2}} (e^{i \sigma p} f(\tau, \sigma),g(\tau, \sigma), e^{-i \sigma p} f(\tau, \sigma),g(\tau, \sigma))\;.
\end{equation} 
In this expression $f(\tau,\sigma)$ and $g(\tau,\sigma)$ are periodic and well--defined functions on a world--sheet. We want to have a closed string, and twists are created for it. Let us consider the asymptotic $x \to \infty$. In this case the monodromy matrix (see \cite{Gromov:2008bz}) gives us some information:
\begin{equation}
 \Omega (x) = P \exp \int d\sigma J_{\sigma}(x) \sim P \exp \int d\sigma j_{\sigma} \sim h^{-1}(2\pi) h(0)
\end{equation} 

Eigenvalues of $\Omega(x)$ are of type 
\begin{equation}
 \{e^{i p_1}, \ldots e^{i p_4} \},
\end{equation} 
where $p_i$ --- the twists, i.e. asymptotics of quasimomenta. In our case computation gives that there will be two non--zero diagonal elements, and the twists are equal to $-p/2$, where $p$ is a magnon momentum on the world--sheet. 

\bibliography{ads_cft} 

\providecommand{\href}[2]{#2}\begingroup\raggedright\begin{thebibliography}{10}

\bibitem{Lipatov:1993yb}
L.~N. Lipatov, ``{High-energy asymptotics of multicolor QCD and exactly
  solvable lattice models},''
\href{http://arxiv.org/abs/hep-th/9311037}{{\tt arXiv:hep-th/9311037}}.

\bibitem{Faddeev:1994zg}
L.~D. Faddeev and G.~P. Korchemsky, ``{High-energy QCD as a completely
  integrable model},''
  \href{http://dx.doi.org/10.1016/0370-2693(94)01363-H}{{\em Phys. Lett.} {\bf
  B342} (1995)  311--322},
\href{http://arxiv.org/abs/hep-th/9404173}{{\tt arXiv:hep-th/9404173}}.

\bibitem{Braun:1998id}
V.~M. Braun, S.~E. Derkachov, and A.~N. Manashov, ``{Integrability of
  three-particle evolution equations in {QCD}},''
  \href{http://dx.doi.org/10.1103/PhysRevLett.81.2020}{{\em Phys. Rev. Lett.}
  {\bf 81} (1998)  2020--2023},
\href{http://arxiv.org/abs/hep-ph/9805225}{{\tt arXiv:hep-ph/9805225}}.

\bibitem{Beisert:2005bm}
N.~Beisert, V.~A. Kazakov, K.~Sakai, and K.~Zarembo, ``{The algebraic curve of
  classical superstrings on AdS(5) x S**5},'' {\em Commun. Math. Phys.} {\bf
  263} (2006)  659--710,
\href{http://arxiv.org/abs/hep-th/0502226}{{\tt arXiv:hep-th/0502226}}.

\bibitem{Minahan:2002ve}
J.~A. Minahan and K.~Zarembo, ``{The Bethe-ansatz for N = 4 super
  Yang-Mills},'' {\em JHEP} {\bf 03} (2003)  013,
\href{http://arxiv.org/abs/hep-th/0212208}{{\tt arXiv:hep-th/0212208}}.

\bibitem{Kazakov:2004qf}
V.~A. Kazakov, A.~Marshakov, J.~A. Minahan, and K.~Zarembo, ``{Classical /
  quantum integrability in AdS/CFT},'' {\em JHEP} {\bf 05} (2004)  024,
\href{http://arxiv.org/abs/hep-th/0402207}{{\tt arXiv:hep-th/0402207}}.

\bibitem{Kazakov:2004nh}
V.~A. Kazakov and K.~Zarembo, ``{Classical / quantum integrability in
  non-compact sector of AdS/CFT},'' {\em JHEP} {\bf 10} (2004)  060,
\href{http://arxiv.org/abs/hep-th/0410105}{{\tt arXiv:hep-th/0410105}}.

\bibitem{Maldacena:1997re}
J.~M. Maldacena, ``{The large N limit of superconformal field theories and
  supergravity},'' {\em Adv. Theor. Math. Phys.} {\bf 2} (1998)  231--252,
\href{http://arxiv.org/abs/hep-th/9711200}{{\tt arXiv:hep-th/9711200}}.

\bibitem{Aharony:2008ug}
O.~Aharony, O.~Bergman, D.~L. Jafferis, and J.~Maldacena, ``{N=6 superconformal
  Chern-Simons-matter theories, M2-branes and their gravity duals},''
\href{http://arxiv.org/abs/0806.1218}{{\tt arXiv:0806.1218 [hep-th]}}.

\bibitem{Stefanski:2008ik}
j.~Stefanski, B., ``{Green-Schwarz action for Type IIA strings on $AdS_4\times
  CP^3$},''
\href{http://arxiv.org/abs/0806.4948}{{\tt arXiv:0806.4948 [hep-th]}}.

\bibitem{Arutyunov:2008if}
G.~Arutyunov and S.~Frolov, ``{Superstrings on AdS4 x CP3 as a Coset
  Sigma-model},''
\href{http://arxiv.org/abs/0806.4940}{{\tt arXiv:0806.4940 [hep-th]}}.

\bibitem{Gromov:2008bz}
N.~Gromov and P.~Vieira, ``{The AdS4/CFT3 algebraic curve},''
\href{http://arxiv.org/abs/0807.0437}{{\tt arXiv:0807.0437 [hep-th]}}.

\bibitem{Minahan:2008hf}
J.~A. Minahan and K.~Zarembo, ``{The Bethe ansatz for superconformal
  Chern-Simons},''
\href{http://arxiv.org/abs/0806.3951}{{\tt arXiv:0806.3951 [hep-th]}}.

\bibitem{Gromov:2008qe}
N.~Gromov and P.~Vieira, ``{The all loop AdS4/CFT3 Bethe ansatz},''
\href{http://arxiv.org/abs/0807.0777}{{\tt arXiv:0807.0777 [hep-th]}}.

\bibitem{Chen:2008qq}
B.~Chen and J.-B. Wu, ``{Semi-classical strings in AdS4*CP3},''
\href{http://arxiv.org/abs/0807.0802}{{\tt arXiv:0807.0802 [hep-th]}}.

\bibitem{Nishioka:2008gz}
T.~Nishioka and T.~Takayanagi, ``{On Type IIA Penrose Limit and N=6
  Chern-Simons Theories},''
\href{http://arxiv.org/abs/0806.3391}{{\tt arXiv:0806.3391 [hep-th]}}.

\bibitem{Bak:2008cp}
D.~Bak and S.-J. Rey, ``{Integrable Spin Chain in Superconformal Chern-Simons
  Theory},''
\href{http://arxiv.org/abs/0807.2063}{{\tt arXiv:0807.2063 [hep-th]}}.

\bibitem{Dorey:2008zy}
N.~Dorey, ``{A Spin Chain from String Theory},''
\href{http://arxiv.org/abs/0805.4387}{{\tt arXiv:0805.4387 [hep-th]}}.

\bibitem{Dorey:2006dq}
N.~Dorey, ``{Magnon bound states and the AdS/CFT correspondence},'' {\em J.
  Phys.} {\bf A39} (2006)  13119--13128,
\href{http://arxiv.org/abs/hep-th/0604175}{{\tt arXiv:hep-th/0604175}}.

\bibitem{Hofman:2006xt}
D.~M. Hofman and J.~M. Maldacena, ``{Giant magnons},'' {\em J. Phys.} {\bf A39}
  (2006)  13095--13118,
\href{http://arxiv.org/abs/hep-th/0604135}{{\tt arXiv:hep-th/0604135}}.

\bibitem{Beisert:2005tm}
N.~Beisert, ``{The su(2|2) dynamic S-matrix},''
\href{http://arxiv.org/abs/hep-th/0511082}{{\tt arXiv:hep-th/0511082}}.

\bibitem{Beisert:2005he}
N.~Beisert and R.~Roiban, ``{The Bethe ansatz for Z(S) orbifolds of N = 4 super
  Yang- Mills theory},'' {\em JHEP} {\bf 11} (2005)  037,
\href{http://arxiv.org/abs/hep-th/0510209}{{\tt arXiv:hep-th/0510209}}.

\bibitem{Grignani:2008te}
G.~Grignani, T.~Harmark, M.~Orselli, and G.~W. Semenoff, ``{Finite size Giant
  Magnons in the string dual of N=6 superconformal Chern-Simons theory},''
\href{http://arxiv.org/abs/0807.0205}{{\tt arXiv:0807.0205 [hep-th]}}.

\bibitem{Gromov:2007aq}
N.~Gromov and P.~Vieira, ``{The AdS(5) x S**5 superstring quantum spectrum from
  the algebraic curve},''
  \href{http://dx.doi.org/10.1016/j.nuclphysb.2007.07.032}{{\em Nucl. Phys.}
  {\bf B789} (2008)  175--208},
\href{http://arxiv.org/abs/hep-th/0703191}{{\tt arXiv:hep-th/0703191}}.

\bibitem{SchaferNameki:2006gk}
S.~Schafer-Nameki, ``{Exact expressions for quantum corrections to spinning
  strings},'' \href{http://dx.doi.org/10.1016/j.physletb.2006.03.033}{{\em
  Phys. Lett.} {\bf B639} (2006)  571--578},
\href{http://arxiv.org/abs/hep-th/0602214}{{\tt arXiv:hep-th/0602214}}.

\bibitem{Gromov:2007cd}
N.~Gromov and P.~Vieira, ``{Constructing the AdS/CFT dressing factor},''
  \href{http://dx.doi.org/10.1016/j.nuclphysb.2007.08.019}{{\em Nucl. Phys.}
  {\bf B790} (2008)  72--88},
\href{http://arxiv.org/abs/hep-th/0703266}{{\tt arXiv:hep-th/0703266}}.

\bibitem{SchaferNameki:2006jf}
S.~Schafer-Nameki and M.~Zamaklar, ``{Stringy effects for spinning strings and
  the Bethe ansatz},''
{\em Fortsch. Phys.} {\bf 54} (2006)  487--495.

\bibitem{Ahn:2008aa}
C.~Ahn and R.~I. Nepomechie, ``{N=6 super Chern-Simons theory S-matrix and
  all-loop Bethe ansatz equations},''
\href{http://arxiv.org/abs/0807.1924}{{\tt arXiv:0807.1924 [hep-th]}}.

\bibitem{Lee:2008ui}
B.-H. Lee, K.~L. Panigrahi, and C.~Park, ``{Spiky Strings on AdS$_4 \times {\bf
  CP}^3$},''
\href{http://arxiv.org/abs/0807.2559}{{\tt arXiv:0807.2559 [hep-th]}}.

\bibitem{Grignani:2008is}
G.~Grignani, T.~Harmark, and M.~Orselli, ``{The SU(2) x SU(2) sector in the
  string dual of N=6 superconformal Chern-Simons theory},''
\href{http://arxiv.org/abs/0806.4959}{{\tt arXiv:0806.4959 [hep-th]}}.

\bibitem{Arutyunov:2006gs}
G.~Arutyunov, S.~Frolov, and M.~Zamaklar, ``{Finite-size effects from giant
  magnons},'' \href{http://dx.doi.org/10.1016/j.nuclphysb.2006.12.026}{{\em
  Nucl. Phys.} {\bf B778} (2007)  1--35},
\href{http://arxiv.org/abs/hep-th/0606126}{{\tt arXiv:hep-th/0606126}}.

\bibitem{Gaiotto:2008cg}
D.~Gaiotto, S.~Giombi, and X.~Yin, ``{Spin Chains in N=6 Superconformal
  Chern-Simons-Matter Theory},''
\href{http://arxiv.org/abs/0806.4589}{{\tt arXiv:0806.4589 [hep-th]}}.

\bibitem{Minahan:2006bd}
J.~A. Minahan, A.~Tirziu, and A.~A. Tseytlin, ``{Infinite spin limit of
  semiclassical string states},'' {\em JHEP} {\bf 08} (2006)  049,
\href{http://arxiv.org/abs/hep-th/0606145}{{\tt arXiv:hep-th/0606145}}.

\bibitem{Gromov:2007ky}
N.~Gromov and P.~Vieira, ``{Complete 1-loop test of AdS/CFT},''
  \href{http://dx.doi.org/10.1088/1126-6708/2008/04/046}{{\em JHEP} {\bf 04}
  (2008)  046},
\href{http://arxiv.org/abs/0709.3487}{{\tt arXiv:0709.3487 [hep-th]}}.

\bibitem{Gromov:2008ie}
N.~Gromov, S.~Schafer-Nameki, and P.~Vieira, ``{Quantum Wrapped Giant
  Magnon},''
\href{http://arxiv.org/abs/0801.3671}{{\tt arXiv:0801.3671 [hep-th]}}.

\end{thebibliography}\endgroup

\end{document}